\pdfoutput=1
\newcommand*{\ATLASLATEXPATH}{}
 
\documentclass[cernpreprint, texlive=2016, UKenglish, LANGEDIT=true, LANGSHOW=false]{\ATLASLATEXPATH atlasdoc}
\LEcontact{Steve Lloyd s.l.lloyd@qmul.ac.uk}

\usepackage{\ATLASLATEXPATH atlaspackage}
\usepackage{\ATLASLATEXPATH atlasbiblatex}
 
\usepackage{\ATLASLATEXPATH atlasphysics}
 
\graphicspath{{logos/}{figures/}}
 
\usepackage{ANA-HIGG-2018-50-PAPER-defs}
 
\usepackage{lscape}
\usepackage{afterpage}
\usepackage{multirow}
\usepackage{comment}
\usepackage{arydshln}

\AtlasTitle{Measurement of $VH$, $H\to b\bar{b}$ production as a function of
the vector-boson transverse momentum in 13~TeV $pp$ collisions with the ATLAS detector}

\AtlasAbstract{
Cross-sections of associated production of a Higgs boson decaying
into bottom-quark pairs and an electroweak gauge boson, $W$ or $Z$,
decaying into leptons are measured as a function of the gauge boson
transverse momentum.
The measurements are performed in kinematic fiducial volumes defined
in the `simplified template cross-section' framework.
The results are obtained using 79.8~fb$^{-1}$ of proton--proton collisions
recorded by the ATLAS detector at the Large Hadron Collider at a
centre-of-mass energy of 13~TeV.
All measurements are found to be in agreement with the Standard Model
predictions, and limits are set on the parameters of an effective Lagrangian
sensitive to modifications of the Higgs boson couplings to the
electroweak gauge bosons.
}
 
\author{The ATLAS Collaboration}
 
\PreprintIdNumber{CERN-EP-2019-019}

\AtlasRefCode{HIGG-2018-50}
 
\PreprintIdNumber{CERN-EP-2019-019}

\AtlasJournalRef{\JHEP 05 (2019) 141}
\AtlasDOI{10.1007/JHEP05(2019)141}

\AtlasCoverSupportingNote{Simplified template cross-section
measurements for the $VH (H\to b\bar{b})$ process with the ATLAS
detector}{https://cds.cern.ch/record/2636121}
 
\AtlasCoverCommentsDeadline{}
 
\AtlasCoverAnalysisTeam{}

\AtlasCoverEgroupEditors{atlas-HIGG-2018-50-editors@cern.ch}
 
\AtlasCoverEgroupEdBoard{atlas-HIGG-2018-50-editorial-board@cern.ch}

\hypersetup{pdftitle={ATLAS document},pdfauthor={The ATLAS Collaboration}}
 
\begin{document}
 
\maketitle

\section{Introduction}
 
A particle
consistent with the Standard Model (SM) predictions for the Higgs
boson~\cite{Englert:1964et, Higgs:1964ia, Higgs:1964pj,
Guralnik:1964eu} was observed in 2012 by the ATLAS and CMS
collaborations~\cite{HIGG-2012-27, CMS-HIG-12-028} at the LHC\@.
Further analysis of ATLAS and CMS data collected in proton--proton
($pp$) collisions at centre-of-mass energies of 7~\TeV, 8~\TeV\ and
13~\TeV\ in two LHC data-taking periods (Runs 1 and 2) has led to
precise measurements of the mass of this particle
(around 125~\GeV)~\cite{HIGG-2014-14, HIGG-2016-33, CMS-HIG-16-041},
tests of its spin and parity ($J^P=0^+$) against alternative
hypotheses~\cite{HIGG-2013-17, CMS-HIG-14-018}, as well as
to measurements of its production and decay
rates~\cite{HIGG-2015-07,CMS-HIG-17-035,HIGG-2018-13}.
 
Recently, experiments at the LHC observed Higgs
boson production in association with weak gauge bosons $V=W,Z$ ($VH$
production)~\cite{HIGG-2018-04} and Higgs boson decays into
pairs of bottom quarks ($H\to b\bar{b}$)~\cite{HIGG-2018-04,
CMS-HIG-18-016}.
With these results, the four most important Higgs boson production modes
predicted by the SM, gluon--gluon fusion (ggF), vector-boson fusion (VBF), and
associated production of a Higgs boson with either a weak gauge boson ($VH$)
or a top-quark pair ($t\bar{t}H$) are established.
Similarly, several of the main modes of Higgs boson decays into
fermionic ($b\bar{b}$, $\tau\tau$)
and bosonic ($WW$, $ZZ$, $\gamma\gamma$) final states are observed.
All results, typically expressed in the form of `signal strengths',
defined as the ratio of the observed to the expected product of the
production cross-section times branching ratio into a certain final
state, are consistent with SM predictions within uncertainties.
 
To probe the kinematic properties of Higgs boson production in more detail,
to reduce the impact of theoretical uncertainties
on the measurements and to make the measurements easier to compare
with future updated calculations, the framework of
simplified template cross-sections (STXS)
has been introduced~\cite{deFlorian:2016spz,Badger:2016bpw}.
In this framework, the cross-sections for the various Higgs boson
production modes are measured in exclusive regions carefully
defined by fiducial selections based
on the kinematic properties of Higgs boson production.
The extrapolation from the phase space selected by the analysis
criteria to that for which the cross-section measurements are
presented is thus reduced.

The STXS measurements are designed
to proceed in stages of increasing granularity with more recorded data.
In `stage 0', cross-sections are measured separately for the four main
production modes
in a fiducial Higgs boson rapidity region
$|y_H|<2.5$,\footnote{ATLAS uses a right-handed coordinate system with its
origin at the nominal interaction point (IP)
in the centre of the detector and the $z$-axis along the beam pipe.
The $x$-axis points from the IP to the centre of the LHC ring,
and the $y$-axis points upwards.
Cylindrical coordinates $(r,\phi)$ are used in the transverse plane,
$\phi$ being the azimuthal angle around the $z$-axis.
The pseudorapidity is defined in terms of the polar angle $\theta$
as $\eta = -\ln \tan(\theta/2)$.
When dealing with massive particles, the rapidity
$y=1/2 \ln[(E+p_z)/(E-p_z)]$ is used,
where $E$ is the energy and $p_z$ is the $z$-component of the momentum.}
mainly driven by the ATLAS and CMS detector acceptances for
most of the reconstructed objects (leptons, photons and $b$-jets).
In `stage 1' these regions are split into 31 subregions according
to kinematic properties such as the number of particle-level jets
with transverse momentum $\pT>30$~\GeV\
(excluding any jets from Higgs boson decays), the transverse momentum of
the Higgs boson, or the transverse momentum
of the weak gauge boson $V$ for $VH$, $V\to$~leptons production.
In simulation, particle-level jets are built by clustering all generated stable particles
($c\tau>10$~mm),
excluding the decay products of the Higgs boson as well as the
neutrinos and charged leptons from the decays of the weak gauge boson,
using the anti-$k_t$ clustering algorithm~\cite{Cacciari:2008gp} with
a radius parameter $R=0.4$.
 
Stage-0 STXS were measured recently with 36.1~\ifb\ of 13~\TeV\ ATLAS data
using $H\to\gamma\gamma$~\cite{HIGG-2016-21}
and $H\to ZZ^*\to 4\ell$ decays~\cite{HIGG-2016-22}, with
results in agreement with SM predictions.
In addition, Refs.~\cite{HIGG-2016-21} and~\cite{HIGG-2016-22}
contain some `reduced' stage-1 STXS measurements of ggF and VBF
regions, after merging together regions where the data lack
sufficient sensitivity to Higgs boson production.
Given the low $VH$ production cross-sections, the only Higgs
boson decay mode that can currently be measured
is $H\to b\bar{b}$, with its large branching ratio of 58\%.
This \type\ presents a measurement of `reduced' stage-1
$VH$ STXS (defined in Section~\ref{sec:selection})
using $H\to b\bar{b}$ decays with 79.8~\ifb\ of 13~\TeV\ $pp$
collisions collected by ATLAS between 2015 and 2017.
The results are used to investigate the strength and tensor
structure of the interactions of the Higgs boson with vector bosons
using an effective Lagrangian approach~\cite{ATL-PHYS-PUB-2017-018}.
 
\section{Data and simulation samples}
\label{sec:detector}
 
The data were collected with the ATLAS detector~\cite{PERF-2007-01,Abbott:2018ikt}
between 2015 and 2017, triggered by isolated
charged leptons or large transverse momentum imbalance, $\ET^\mathrm{miss}$.
Only events with good data quality were kept.
 
The Monte Carlo simulation samples used for the measurements presented here
are identical to those used for the measurement of the inclusive
$VH$, $H\to b\bar{b}$ signal strength~\cite{HIGG-2018-04}.
Several samples of simulated events were produced for the
signal ($q\bar{q} \to WH$, $q\bar{q}\to ZH$
and $gg\to ZH$) and main background
($t\bar{t}$, single-top, $V$+jets and diboson) processes.
They were used to optimise the analysis criteria and to
determine the expected signal and background distributions of the
discriminating variables used in the final fit to the data.
The multijet background is largely suppressed by the selection
criteria and is estimated using data-driven techniques.
 
The signal templates in each STXS region were obtained from simulated
$q\bar{q} \to WH$  and $q\bar{q}\to ZH$ events with zero or
one additional jet, calculated at next-to-leading order (NLO),
generated with the \POWHEGBOXV{v2} + \textsc{GoSam} + \textsc{MiNLO}
generators~\cite{Alioli:2010xd,arXiv:1111.2034,arXiv:1206.3572,Luisoni:2013kna}.
The contribution from loop-induced $gg \to ZH$ production was
simulated at leading order (LO) using the \POWHEGBOXV{v2}
generator~\cite{Alioli:2010xd}.
Additional scale factors were applied to the $q\bar{q} \to V H$
processes as a function of the generated
vector-boson transverse momentum ($\pt^{V}$) to
account for electroweak (EW) corrections at NLO\@.
These factors were determined from the ratio between the $V H$ differential
cross-sections computed with and without these corrections by the \textsc{Hawk} program~\cite{Denner:2011id,Denner:2014cla}.
The mass of the Higgs boson was fixed at 125~\GeV.
 
In the measurement of the $pp \to  ZH$ cross-sections, the relative
contributions of the $q\bar{q} \to Z H$ and $gg \to ZH$ processes
are determined by the most accurate theoretical cross-section
predictions currently available: next-to-next-to-leading order (NNLO)
in QCD and NLO in EW~\cite{Ciccolini:2003jy,Brein:2003wg,Ferrera:2011bk,Brein:2011vx,Ferrera:2013yga,Ferrera:2014lca,Campbell:2016jau}
for $q\bar{q} \to Z H$, and next-to-leading order and
next-to-leading logarithm (NLO+NLL) in QCD~\cite{Altenkamp:2012sx,Hespel:2015zea,ggzhnll,Harlander:2013mla,Brein:2012ne}
for $gg \to ZH $.
 
\section{Event selection and categorisation}
\label{sec:selection}
 
The object reconstruction, event selection and classification
into categories used for the measurements, are identical to those described in
Ref.~\cite{HIGG-2018-04}.
The selection and the event categories are briefly summarised below.
 
Events are retained if they are consistent with one of the typical
signatures of $VH$, $H\to b\bar{b}$ production and decay, with
$Z\to \nu\bar{\nu}$, $W\to \ell\nu$  or
$Z\to \ell\ell$ ($\ell=e,\mu$).
Vector-boson decays into $\tau$-leptons are not
targeted explicitly. However, they satisfy the selection criteria with
reduced efficiency in the case of leptonic $\tau$-lepton decays.
 
In particular, events are kept if they contain
at most two isolated electrons or muons,
and two good-quality high-$\pT$ ($>45, 20$~\GeV) jets with $|\eta|<2.5$
satisfying $b$-jet identification ('$b$-tagging') requirements
(which have an average
efficiency of 70\% for jets containing $b$-hadrons that are
produced in inclusive $t\bar{t}$ events~\cite{PERF-2016-05}).
The two $b$-jet candidates are used to reconstruct
the Higgs boson candidate; their invariant mass is denoted by $m_{bb}$.
Additional jets are required to have $\pT > 20$~\GeV\ for $|\eta|<2.5$
or $\pT > 30$~\GeV\ for $2.5<|\eta|<4.5$, and not be
identified as $b$-jets.
 
Events with either zero, one or two isolated electrons or muons
are classified as `0-lepton', `1-lepton' or `2-lepton'
events, respectively.
The 0-lepton events and the 1-lepton events are required to have
transverse momentum imbalance, as expected from the neutrinos from
$Z\to\nu\bar{\nu}$ or $W\to\ell\nu$ decays; in the 2-lepton events,
the leptons must have the same flavour (and opposite charge for
events with muons) and an invariant mass close to the $Z$ boson mass.
 
Additional requirements are applied to suppress background
from QCD production of multijet events in the 0-lepton and 1-lepton
channels.
To suppress the large $t\bar{t}$ background, events with
four or more jets are discarded in the 0-lepton and
1-lepton channels.
Finally, a requirement on the reconstructed transverse momentum
$\pT^{V,\mathrm{r}}$ of the vector boson $V$ is applied.
It is  computed, depending on the number, $N_\mathrm{lep}$,
of selected electrons and muons, as either
the missing transverse momentum $\ET^\mathrm{miss}$ ($N_\mathrm{lep}=0$),
the magnitude of the vector sum of the missing transverse momentum
and the lepton \pT\ ($N_\mathrm{lep}=1$),
or the dilepton \pT\ ($N_\mathrm{lep}=2$).
The minimum value of $\pT^{V,\mathrm{r}}$ is 150~\GeV\ in the 0- and 1-lepton
channels, and 75~\GeV\ in the 2-lepton channel.
 
Events satisfying the previous criteria are classified into eight
categories (also called signal regions in the following),
shown in Table~\ref{tab:anaCat}, with different
signal-to-background ratios. These categories are defined
by the number of jets, $N_\mathrm{jet}$
(including the two $b$-jet candidates),
$N_\mathrm{lep}$, and $\pT^{V,\mathrm{r}}$.
Additional categories (also called control regions in the following)
containing events satisfying alternative selections
are introduced to constrain some background processes
such as $W$ boson production in association with jets
containing heavy-flavour hadrons ($W$+HF), or top-quark pair
production.
The signal contribution in such categories is expected to be negligible.
\begin{table}[!htbp]
\centering
\caption{Summary of the reconstructed-event categories.
Categories with relatively large fractions of the total
expected signal yields are referred to as `signal regions' (SR),
while those with negligible expected signal yield,
mainly designed to constrain some background processes,
are called `control regions' (CR).
The quantity $\mtop$ is the reconstructed mass of a
semileptonically decaying top-quark candidate in the
1-lepton channel.
The calculation of $\mtop$ uses the four-momenta of one of the two
$b$-jet candidates, the lepton, and the hypothetical neutrino produced
in the event. The neutrino four-momentum is derived using the $W$
boson mass constraint~\cite{HIGG-2018-04} and $\mtop$
is then reconstructed from the combination of the $b$-jet candidate
and the value of the neutrino longitudinal momentum that yields the
smallest top-quark candidate mass.
The $\mtop\leq225$~\GeV\ requirement in the 1-lepton signal region
is needed to maintain orthogonality with the $W$+HF control region.
}
\label{tab:anaCat}
\setlength{\extrarowheight}{2pt}
\begin{tabular}{ l | c |  c | c | c | c}
\hline
\hline
\multirow{3}{*}{Channel}                    & \multicolumn{5}{c}{Categories} \\
\cline{2-6}
& \multicolumn{2}{c}{$75$~\GeV~$<\pT^{V,\mathrm{r}}<150$~\GeV} & \multicolumn{3}{|c}{$\pT^{V,\mathrm{r}}>150$~\GeV} \\
\cline{2-6}
& $2$ jets       & $\geq 3$ jets       & $2$ jets     & $3$ jets     & $\geq 3$ jets   \\
\hline
\hline
0-lepton                                    & $-$            & $-$            & SR          & SR          & $-$   \\
\hline
1-lepton & \multicolumn{5}{c}{}   \\
$~~\mbb\geq75$~\GeV\ or $\mtop\leq225$~\GeV\    & $-$            & $-$            & SR          & SR          & $-$   \\
$~~\mbb<75$~\GeV\ and $\mtop>225$~\GeV\         & $-$            & $-$            & CR          & CR          & $-$   \\
\hline
2-lepton & \multicolumn{5}{l}{}   \\
$~~ee$ and $\mu\mu$ channels                & SR             & SR             & SR          & $-$         & SR    \\
$~~e\mu$ channel                            & CR             & CR             & CR          & $-$         & CR    \\
\hline
\hline
\end{tabular}
\end{table}
 
 
\section{Cross-section measurements}
The reduced $VH$, $V\to$~leptons stage-1 STXS regions used in this
\type\ are summarised in Table~\ref{tab:STXS_reducedstage1},
which also indicates which reconstructed-event
categories are most sensitive in each region.
All leptonic decays of the weak gauge bosons (including $Z\to\tau\tau$ and
$W\to\tau\nu$) are considered for the STXS definition.
 
Compared to the original stage-1 proposal presented in Ref.~\cite{deFlorian:2016spz},
the following changes have been made for the reduced $VH$, $V\to$~leptons
stage-1 STXS regions of Table~\ref{tab:STXS_reducedstage1}:
\begin{itemize}
\item
the $\pT^Z<150~\GeV$ stage-1 regions are split into two
subregions, $\pT^Z<75~\GeV$ and $75<\pT^Z<150$~\GeV,
to avoid theory uncertainties from extrapolations to
a phase space not accessible to this measurement;
\item
an additional $gg\to ZH$, $\pT^Z~>250$~\GeV\ region
has been introduced, similarly to what is already done
for $q\bar{q}\to ZH$.
\end{itemize}
These two changes lead to a total of 14 modified stage-1 regions,
which are then combined together in reduced stage-1 regions, chosen
to keep the total uncertainty in the measurements near or below 100\%,
in the following way:
\begin{itemize}
\item
the $q\bar{q}\to ZH$ and $gg\to ZH$ regions are merged.
There are currently not enough data events to distinguish
$q\bar{q}\to ZH$ from gluon-induced $ZH$ production
despite their different kinematic properties;
\item
the $150<\pT^V<250$~\GeV\ regions with zero or at least one
particle-level jet are merged.
\end{itemize}
 
Two sets of reduced stage-1 regions are considered.
In one, called the `5-POI (parameters of interest)'
scheme, five cross-sections, three for $ZH$ production ($75<\pT^Z<150$~\GeV,
$150<\pT^Z<250$~\GeV\ and $\pT^Z>250$~\GeV) and two for
$WH$ production ($150<\pT^W<250$~\GeV\ and $\pT^W>250$~\GeV), are measured.
In the other one, called the  `3-POI' scheme,
three cross-sections, two for $ZH$ ($75<\pT^Z<150$~\GeV\
and $\pT^Z>150$~\GeV) and one for $WH$ ($\pT^W>150$~\GeV),
are measured.
The 5-POI scheme leads to measurements that have total
uncertainties larger than those in the 3-POI scheme, but are more
sensitive to enhancements at high $\pT^V$ from potential anomalous
interactions between the Higgs boson and the EW gauge bosons.
 
\begin{table}[!htbp]
\caption{
The 3-POI and 5-POI `reduced stage-1' sets
of merged regions used for the measurements, the corresponding
kinematic regions of the stage-1 $VH$ simplified template cross-sections,
and the reconstructed-event categories that are most sensitive in each
merged region.
The stage-1 regions are modified (i) by splitting the two
$ZH$, $\pT^Z<150~\GeV$ regions (from $q\bar{q}$ and $gg$)
into four regions, based on whether $\pT^Z<75~\GeV$ or
$75<\pT^Z<150~\GeV$; (ii) by adding a $\pT^Z<250$~\GeV\ requirement
to the $gg\to ZH$, $\pT^Z>150$~\GeV\ regions (with zero or at least one
extra particle-level jet), and (iii) by adding a separate $gg\to ZH,
\pT^Z>250$~\GeV\ region.
The three regions $WH$, $\pT^W<150$~\GeV,
$q\bar{q}\to ZH$, $\pT^Z<75$~\GeV\
and $gg\to ZH$, $\pT^Z<75$~\GeV,
in which the current analysis is not sensitive and
whose corresponding cross-sections are fixed to the SM prediction
in the fit, are not shown.
}
\label{tab:STXS_reducedstage1}
\begin{center}
\resizebox{\textwidth}{!}{
\setlength{\extrarowheight}{4pt}
\begin{tabular}{lll|lll}
\hline \hline
Merged region                                       & Merged region                                        & \multirow{3}{*}{Stage 1 (modified) STXS region} & \multicolumn{3}{l}{Reconstructed-event categories} \\
3-POI scheme                                        & 5-POI scheme                                         &                                                 & \multicolumn{3}{l}{with largest sensitivity} \\
&                                                      &                                                 & $N_\mathrm{lep}$ & $\pT^{V,\mathrm{r}}$ interval & $N_\mathrm{jet}$ \\
\hline
\hline
\multirow{3}{*}{$WH$, $\pT^W>150~\GeV$}             & \multirow{2}{*}{$WH$, $150<\pT^W<250~\GeV$}          & $q\bar{q}\to WH$, $150 <\pT^W<250~\GeV$, 0-jet        & \multirow{3}{*}{1} & \multirow{3}{*}{$>150$~\GeV} & \multirow{3}{*}{2, 3} \\
&                                                      & $q\bar{q}\to WH$, $150 <\pT^W<250~\GeV$, $\geq 1$-jet & & & \\
\cline{2-3}
& $WH$, $\pT^W>250~\GeV$                               & $q\bar{q}\to WH$, $\pT^W>250~\GeV$                    & & & \\
\hline
\multirow{2}{*}{$ZH$, $75<\pT^Z<150~\GeV$} & \multirow{2}{*}{$ZH$, $75<\pT^Z<150~\GeV$}                    & $q\bar{q}\to ZH$, $75<\pT^Z<150~\GeV$ & \multirow{2}{*}{2} & \multirow{2}{*}{75--150~\GeV} & \multirow{2}{*}{2, $\geq 3$} \\
&                                                      & $gg\to ZH$, $75<\pT^Z<150~\GeV$       & & & \\
\hline
\multirow{6}{*}{$ZH$, $\pT^Z>150~\GeV$}             & \multirow{4}{*}{$ZH$, $150<\pT^Z<250~\GeV$}          & $q\bar{q}\to ZH$, $150<\pT^Z<250~\GeV$, 0-jet & & & \\
&                                                      & $gg\to ZH$, $150<\pT^Z<250~\GeV$, 0-jet       &  & & \\
&                                                      & $q\bar{q}\to ZH$, $150<\pT^Z<250~\GeV$, $\geq 1$-jet & {0} & {$>150$~\GeV} & {2, 3}  \\
&                                                      & $gg\to ZH$, $150<\pT^Z<250~\GeV$, $\ge 1$-jet & {2} & {$>150$~\GeV} & {2, $\geq$3}  \\
\cline{2-3}
& \multirow{2}{*}{$ZH$, $\pT^Z>250~\GeV$}              & $q\bar{q}\to ZH$, $\pT^Z>250~\GeV$            & & & \\
&                                                      & $gg\to ZH$, $\pT^Z>250$~\GeV\                   & & & \\
\hline \hline
\end{tabular}
}
\end{center}
\end{table}
 
The reconstructed-event categories do not distinguish between events
with generated $\pT^V$ below or above 250~\GeV.
Discrimination between the two $\pT^V$ regions 150--250~\GeV\ and
$>250$~\GeV\ is provided by the different shapes
of the boosted-decision-tree discriminant ($\mathrm{BDT}_{VH}$)
used in the final fit to the data,
as illustrated in Figure~\ref{fig:mva1Lep} in the case of the
1-lepton, 2-jet category.
This arises from the fact that the reconstructed $\pT^{V,\mathrm{r}}$
is largely correlated with the $\mathrm{BDT}_{VH}$ output, for which it
constitutes one of the most discriminating input variables
together with $m_{bb}$ and the angular separation of the two $b$-jets.
 
\begin{figure}[!htbp]
\begin{center}
\includegraphics[width=0.7\linewidth]{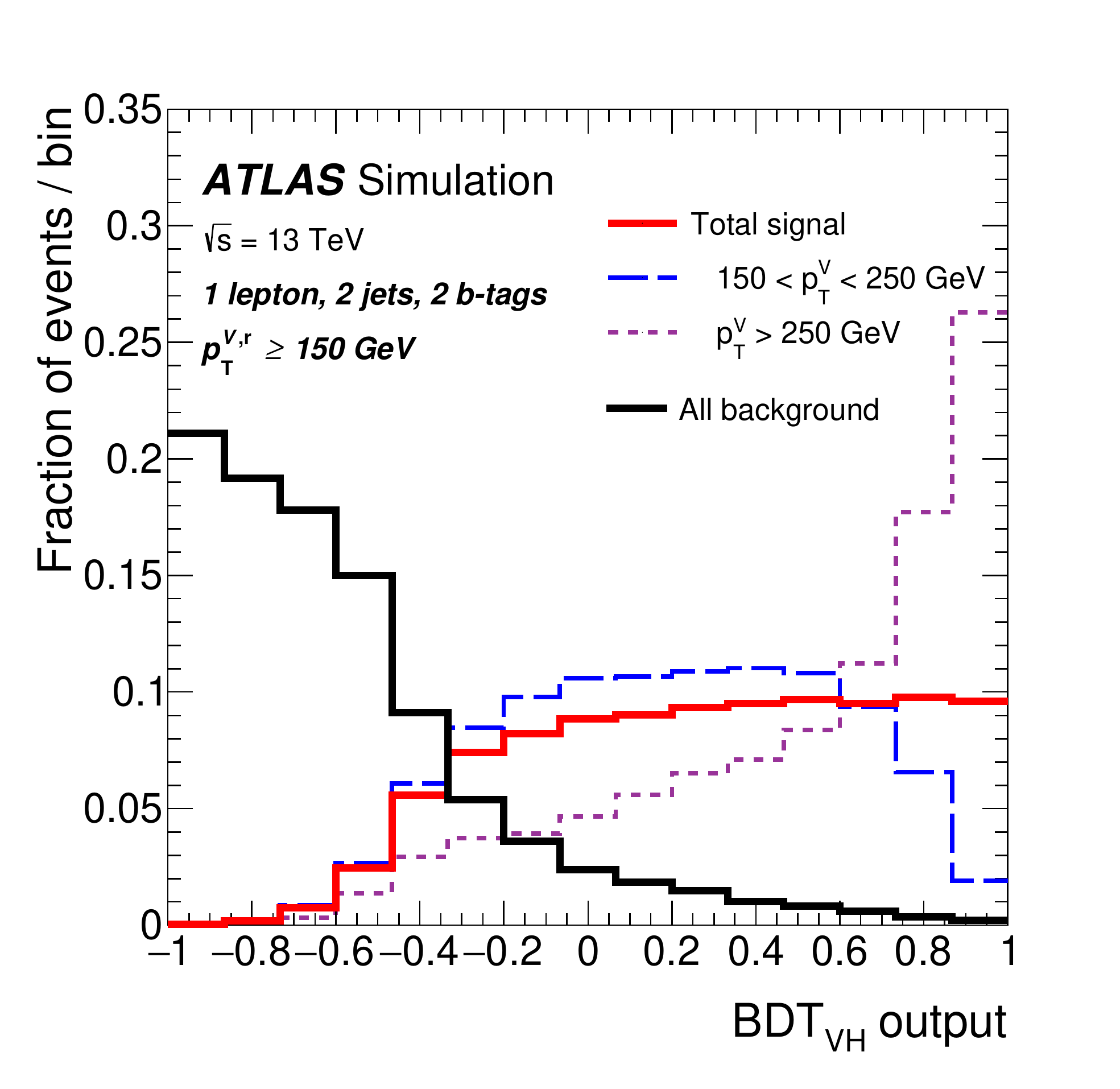}
\caption{$\mathrm{BDT}_{VH}$ distributions for different $\pT^V$ STXS
regions in the 1-lepton, 2-jet reconstructed-event category.
Only regions contributing at least 10\% of the expected signal
yield in the reconstructed-event category are displayed.
The distributions of the total signal and background are also shown.
The $\mathrm{BDT}_{VH}$ distributions are scaled to the same
(unit) area to highlight the shape differences.
}
\label{fig:mva1Lep}
\end{center}
\end{figure}
 
The product of the signal cross-section times the $H\to b\bar{b}$
branching ratio and the total leptonic decay branching ratio for $W$
or $Z$ bosons is determined in each of the reduced stage-1 regions
by a binned maximum-likelihood fit to the data.
The cross-sections are not constrained to be positive in the fit.
Signal and background templates of the discriminating variables,
determined from the simulation or data control regions,
are used to extract the signal and background yields.
A simultaneous fit is performed to all the signal and control regions.
Systematic uncertainties are included in the likelihood function as
nuisance parameters.
 
The likelihood function is very similar to that described
in Ref.~\cite{HIGG-2018-04}.
In particular, the same observables are used, namely
$\mathrm{BDT}_{VH}$
in the signal regions and either the invariant mass \mbb\
of the two $b$-jets or the event yield in the control regions.
The treatment of the background and of its uncertainties
is also unchanged.
The only differences relative to the likelihood function
in Ref.~\cite{HIGG-2018-04}
concern the treatment of the signal:
\begin{itemize}
\item Instead of a single signal shape (for $\mathrm{BDT}_{VH}$ or \mbb)
or yield per category, multiple shapes or yields are introduced, one
for each reduced stage-1 STXS region under study.
\item Instead of a single parameter of interest, the inclusive signal
strength,
the fit has multiple parameters of interest,
i.e. the cross-sections of the reduced stage-1
regions, multiplied by the $H\to b\bar{b}$ and $V\to$~leptons
branching ratios.
\item Overall theoretical cross-section and branching ratio uncertainties,
which affect the signal strength measurements but not the
STXS measurements, are not included in the likelihood function.
\end{itemize}
 
The expected signal shapes of the discriminating variable distributions
and the acceptance times efficiency (referred to as `acceptance'
in the following)
in each reduced stage-1 region are determined from simulated samples
of SM $VH$, $V\to$~leptons, $H\to b\bar{b}$ events.
The acceptance of each reconstructed-event category for signal events
from the different regions of the 5-POI reduced stage-1 scheme
is shown in Figure~\ref{fig:5POI_acceptance}.
The fraction of signal events in each reconstructed-event
category originating from the different regions in the same scheme
is shown in Figure~\ref{fig:5POI_purity}.

\begin{figure}[!htbp]
\begin{center}
\subfloat[]{\includegraphics[width=0.7\linewidth]{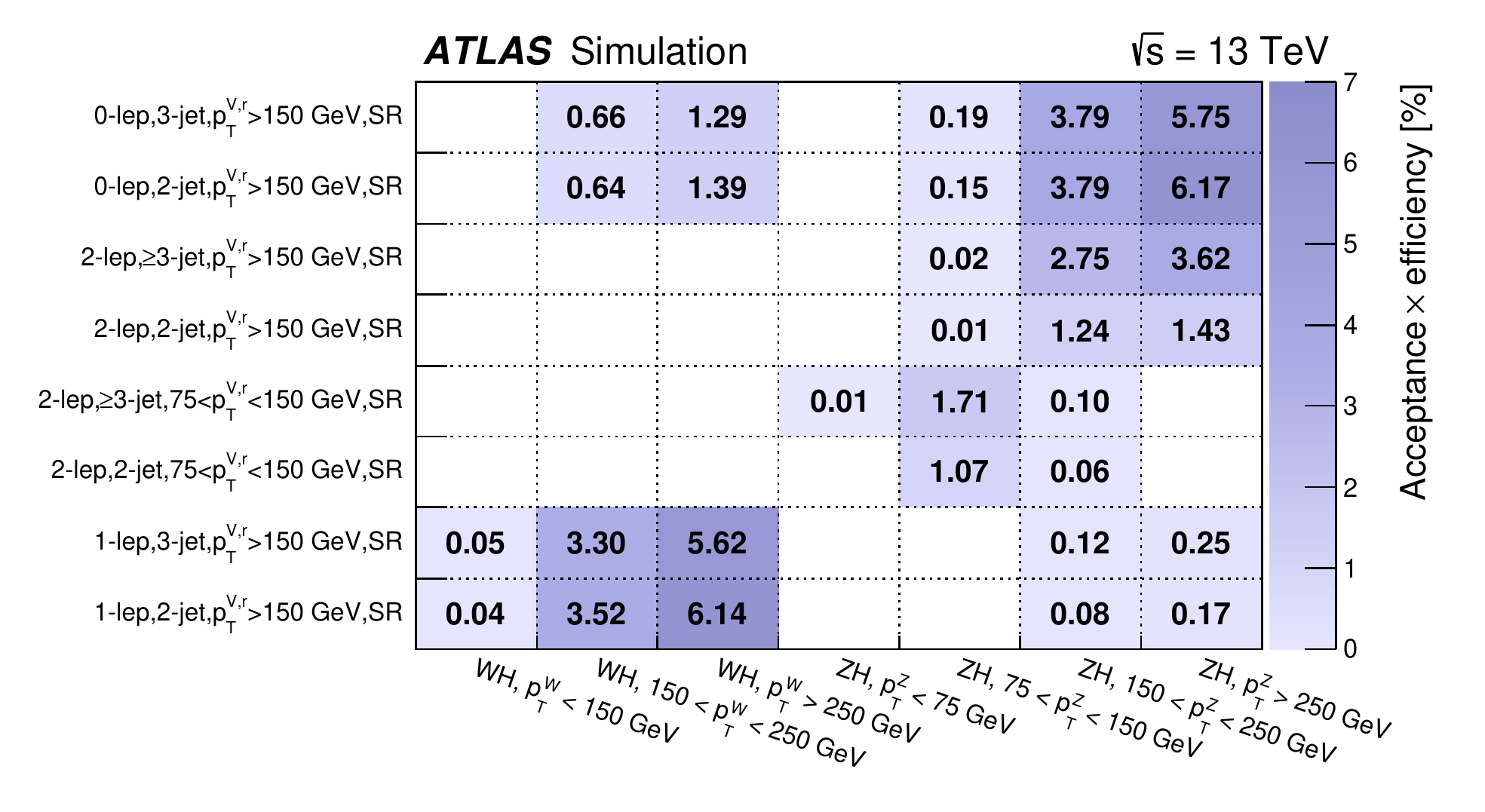}\label{fig:5POI_acceptance}} \\
\subfloat[]{\includegraphics[width=0.7\linewidth]{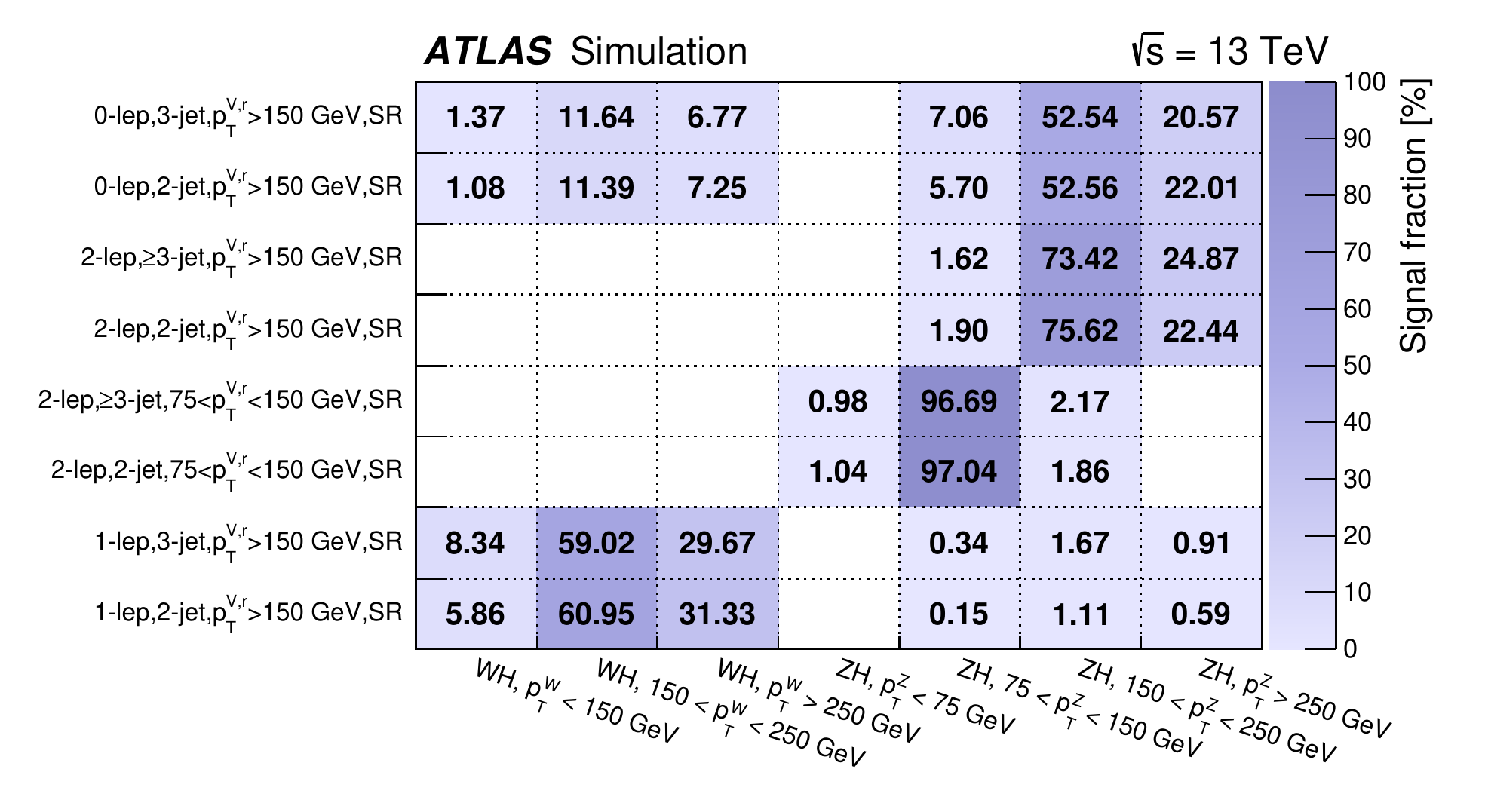}\label{fig:5POI_purity}}
\caption{In the 5-POI reduced stage-1 scheme,
(a) the acceptance (including the efficiency of the experimental
selection) for $VH$, $V\to$~leptons, $H\to b\bar{b}$ events
of each reconstructed-event category ($y$-axis) for
each STXS signal region ($x$-axis), in percent;
(b) the fraction of signal (in percent) from each STXS signal region
($x$-axis) in every reconstructed-event category ($y$-axis).
Entries with acceptance times efficiency below 0.01\% or signal
fractions below 0.1\% are not shown.
}
\end{center}
\end{figure}
 
As shown in Figure~\ref{fig:5POI_acceptance}, the
current analysis is not sensitive to $WH$ events with
$\pT^W<150$~\GeV\ and to $ZH$ events
with $\pT^Z<75$~\GeV, since their acceptance in each
category is at the level of 0.1\% or smaller.
Therefore, in the fits the signal cross-section in these regions is
constrained to the SM prediction, within the theoretical
uncertainties.
Since these regions contribute only marginally to the
selected event sample, the impact on the final results is negligible.
A cross-check in which the relative signal cross-section uncertainty for
the $\pT^W<150$~\GeV\ and $\pT^Z<75$~\GeV\ regions is conservatively set to
70\% of the prediction (i.e. about seven times the nominal uncertainty)
leads to variations of the measured STXS below 1\%.
 
The sources of systematic uncertainty are identical to those described
in Ref.~\cite{HIGG-2018-04}, except for those
associated with the Higgs boson signal simulation, which are
re-evaluated~\cite{ATL-PHYS-PUB-2018-035}.
In this re-evaluation the uncertainties are separated into two groups:
\begin{itemize}
\item uncertainties affecting signal modelling --
i.e.~acceptance and shape of kinematic distributions -- in each of the
three or five reduced stage-1 regions (hereafter referred to as
theoretical modelling uncertainties), and
\item uncertainties in the
prediction of the production cross-section for each of these regions
(hereafter referred to as theoretical cross-section uncertainties).
\end{itemize}
While theoretical modelling uncertainties enter the measurement of the
STXS, theoretical cross-section uncertainties do not affect the results,
but only the predictions with which they are compared.
The consequent reduction of the impact of the theoretical uncertainties
on the results with respect to the signal strength measurements is one
of the main advantages of measuring STXS.
 
The two groups of systematic uncertainties are estimated for
high-granularity STXS regions, and then merged into the reduced
scheme under consideration. This approach makes it easy to compute
the systematic uncertainties for merging schemes different from those
presented here.
The  uncertainties are evaluated by dividing the phase space
into five $\pT^V$ regions (with the following
lower edges: 0~\GeV, 75~\GeV, 150~\GeV, 250~\GeV\ and 400~\GeV), and each
$\pT^V$ region into three bins depending on the number of
particle-level jets (zero, one, or at least two), independently for
the $q\bar{q}\to VH$ and $gg\to ZH$ processes.
When two STXS regions are merged, their relative theoretical
cross-section uncertainties lead to a modelling uncertainty.
These uncertainties
are evaluated as the remnant of the theoretical cross-section
uncertainties for the high-granularity regions after the subtraction
of the theoretical cross-section uncertainty for the merged region.
 
The high-granularity regions are used to calculate theoretical
cross-section uncertainties for the missing higher-order terms in the QCD
perturbative expansion and for the uncertainties induced by the choices
of the parton distribution function (PDF) and $\alphas$.
Fourteen independent sources of uncertainties due to the missing higher-order
terms lead to total uncertainties of 3\%--4\% for $q\bar{q}\to VH$ and
40\%--50\% for $gg\to ZH$  with $\pT^V>$ 75 \GeV~\cite{ATL-PHYS-PUB-2018-035}.
Thirty-one independent sources of PDF and $\alphas$ uncertainties, each
of them usually smaller than 1\%, lead to a total quadrature sum
between 2\% and 3\% depending on the STXS region.
The theoretical modelling uncertainties change the shapes of the
reconstructed $\pt^{V,\mathrm{r}}$ and $m_{bb}$ distributions in the same way
as described in Ref.~\cite{HIGG-2018-04}. Four independent sources for the QCD
expansion and two independent sources for the PDF and $\alphas$ choices
are considered.
 
Systematic uncertainties in the signal acceptance and shape of
the $\pT^{V,\mathrm{r}}$ and $m_{bb}$ distributions due to the parton shower (PS)
and underlying event (UE) models are estimated from the variations of
acceptance and shapes of simulated events after changing the
\PYTHIAV{8} PS parameters or after replacing \PYTHIAV{8} with
\HERWIGV{7} for the PS and UE models~\cite{HIGG-2018-04}.
The signal acceptance uncertainties due to the PS and UE models
(five independent sources) are typically of the order of 1\% (5\%--15\%)
with a maximum of 10\% (30\%) for the $q\bar{q}\to VH$  ($gg\to ZH$)
production mode. Two independent nuisance parameters account for
the systematic uncertainties induced by the PS and UE models
in the $\pT^{V,\mathrm{r}}$ and $m_{bb}$ distributions.
In addition, a systematic uncertainty due to the EW
corrections is parameterised as a change in shape of the $\pT^V$
distributions for the $q\bar{q}\to VH$ processes~\cite{HIGG-2018-04}.
 
\section{Results}
 
The measured reduced stage-1 $VH$ cross-sections times
the $H\to b\bar{b}$ and $V\to$~leptons branching ratios,
$\sigma\times B$, in the 5-POI and 3-POI schemes,
together with the SM predictions, are summarised in
Table~\ref{tab:XS3and5POI}.
The results of the 5-POI scheme are also illustrated in Figure~\ref{fig:Xsections}.
The SM predictions are shown together with the theoretical cross-section
uncertainty for the merged regions computed as described in the
previous section.
The measurements are in agreement with the SM predictions.
 
The cross-sections measured in the $\pT^V>150$~\GeV\ intervals are not
equal to the sum of those measured for $150<\pT^V<250$~\GeV\ and
$\pT^V>250$~\GeV. This is because the
signal template  for $\pT^V>150$~\GeV\ in the 3-POI fit is computed
from the sum of
the templates of the two regions assuming that the ratio of yields
in those regions is that predicted by the SM, while in the 5-POI fit
the normalisations of the two templates are floated independently.
 
The cross-sections are measured with relative uncertainties varying
between 50\% and 125\% in the 5-POI case, and between 29\% and 56\%
for the 3-POI\@.
The largest uncertainties are statistical, except for the $WH$
cross-sections with
$\pT^W>150$~\GeV\ in the 3-POI case and with $150<\pT^W<250$~\GeV\ in
the 5-POI case.
In the 5-POI case, an anti-correlation of the order of 40\%--60\% is
observed between the cross-sections in the ranges $\pT^V>250$~\GeV\ and
$150<\pT^V<250$~\GeV, which are measured with the same reconstructed-event
categories.
 
The dominant systematic uncertainties are due to the
limited number of simulated background events and the
theoretical modelling of the background processes.
The uncertainties due to the theoretical modelling of the $VH$ signal
are small, with relative values ranging between 6\% and 12\%.
The uncertainties in the predictions are 2--3 times larger for $ZH$
than for $WH$ in the same $\pT^V$ interval due to the limited precision
of the theoretical calculations of the $gg\to ZH$ process.
 
\begin{table}[!htbp]
\setlength{\extrarowheight}{4pt}
\begin{center}
\caption{Best-fit values and uncertainties for the $VH$,
$V\to$~leptons reduced stage-1 simplified template cross-sections
times the $H\to b\bar{b}$ branching ratio, in the 5-POI (top five rows)
and 3-POI (bottom three rows) schemes.
The SM predictions for each region, computed using the inclusive
cross-section
calculations and the simulated event samples described in
Section~\ref{sec:detector}, are also shown.
The contributions to the total uncertainty in the measurements from
statistical (Stat.~unc.) or systematic uncertainties (Syst.~unc.) in
the signal modelling (Th.~sig.), background modelling (Th.~bkg.),
and in experimental performance (Exp.) are given separately.
The total systematic uncertainty, equal to the difference in quadrature between
the total uncertainty and the statistical uncertainty, differs from the
sum in quadrature of the Th. Sig., Th. Bkg., and Exp. systematic
uncertainties due to correlations.
All leptonic decays of the $V$ bosons (including those to $\tau$-leptons, $\ell=e,\mu,\tau$) are considered.}
\label{tab:XS3and5POI}
\resizebox{\textwidth}{!}{
\begin{tabular}{l | rcr | rcr | cr | cr | cr | cr }
\hline \hline
Measurement region                                      & \multicolumn{3}{c|}{SM prediction}  & \multicolumn{3}{c|}{Result} & \multicolumn{2}{c}{Stat. unc.}    & \multicolumn{6}{c}{Syst. unc. [fb]}                     \\
\cline{10-15}
($|y_H|<2.5$, $H\to b\bar{b}$)                          & \multicolumn{3}{c|}{[fb]}           & \multicolumn{3}{c|}{[fb]}   & \multicolumn{2}{c}{[fb]}           & \multicolumn{2}{c}{Th. sig.}           & \multicolumn{2}{c}{Th. bkg.}            & \multicolumn{2}{c}{Exp.}            \\
\hline\hline
\multicolumn{15}{c}{5-POI scheme} \\
\hline
$W\rightarrow \ell\nu$; $150<p_{\text{T}}^{V}<250$~\GeV\         & $24.0$ & $\pm$ & $1.1$        & $20$ & $\pm$ & $25$    &  $\pm$ & $17$   & $\pm$ & 2  & $\pm$ & 13 & $\pm$ & 9 \\
$W\rightarrow \ell\nu$; $p_{\text{T}}^{V}>250$~\GeV\             & $7.1$  & $\pm$ & $0.3$        & $ 8.8$ & $\pm$ & $ 5.2 $   &  $\pm$ & $4.4$    & $\pm$ & 0.5  & $\pm$ & 2.5  & $\pm$ & 0.9 \\
$Z\rightarrow \ell\ell,\nu\nu$; $75<p_{\text{T}}^{V}<150$~\GeV\     & $50.6$ & $\pm$ & $4.1$        & $81$ & $\pm$ & $ 45$   &  $\pm$ & $35$   & $\pm$ & 10 & $\pm$ & 21 & $\pm$ & 19 \\
$Z\rightarrow \ell\ell,\nu\nu$; $150<p_{\text{T}}^{V}<250$~\GeV\    & $18.8$ & $\pm$ & $2.4$        & $14$ & $\pm$ & $ 13$   &  $\pm$ & $11$   & $\pm$ & 1  & $\pm$ & 6  & $\pm$ & 3 \\
$Z\rightarrow \ell\ell,\nu\nu$; $p_{\text{T}}^{V}>250$~\GeV\        & $4.9$  & $\pm$ & $0.5$        & $ 8.5$ & $\pm$ & $ 4.0 $   &  $\pm$ & $3.7$    & $\pm$ & 0.8  & $\pm$ & 1.2  & $\pm$ & 0.6 \\
\hline
\multicolumn{15}{c}{3-POI scheme} \\
\hline
$W\rightarrow \ell\nu$; $p_{\text{T}}^{V}>150$~\GeV\             & $31.1$ & $\pm$ & $1.4$           & $35$ & $\pm$ & $14$          & $\pm$ & 9            & $\pm$ & 2 & $\pm$ & 9   & $\pm$ & 4 \\
$Z\rightarrow \ell\ell,\nu\nu$; $75<p_{\text{T}}^{V}<150$~\GeV\     & $50.6$ & $\pm$ & $4.1$           & $81$ & $\pm$ & $45$          & $\pm$ & 35           & $\pm$ & 10& $\pm$ & 21  & $\pm$ & 19 \\
$Z\rightarrow \ell\ell,\nu\nu$; $p_{\text{T}}^{V}>150$~\GeV\        & $23.7$ & $\pm$ & $3.0$           & $28.4$ & $\pm$ & $8.1$           & $\pm$ & 6.4            & $\pm$ & 2.4 & $\pm$ & 3.6   & $\pm$ & 2.3 \\
\hline \hline
\end{tabular}
}
\end{center}
\end{table}

\begin{figure}[!htbp]
\begin{center}
\includegraphics[width=0.7\textwidth]{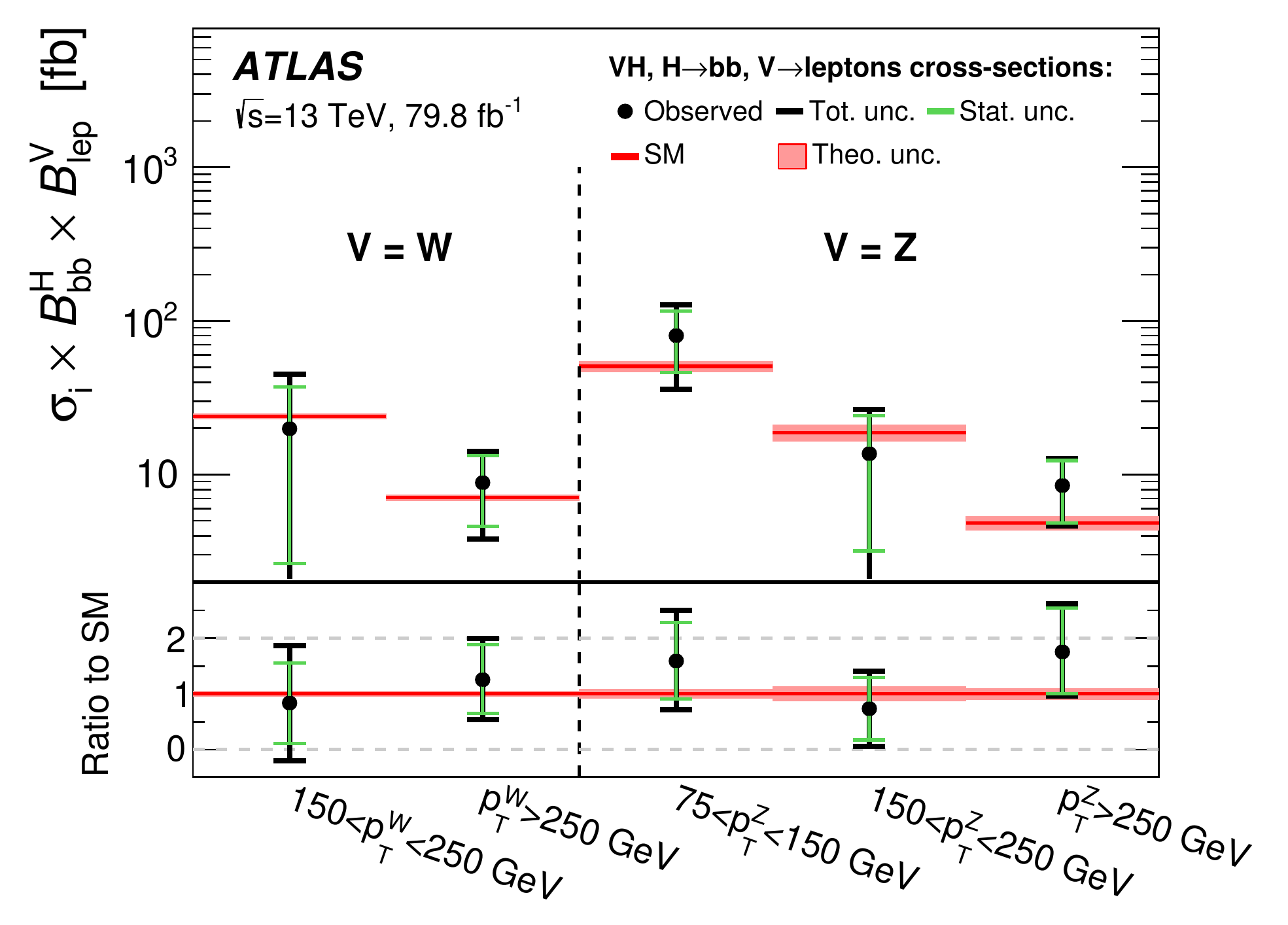}
\caption{Measured $VH$, $V\to$~leptons reduced stage-1 simplified
template cross-sections times the $H\to b\bar{b}$ branching ratio.}
\label{fig:Xsections}
\end{center}
\end{figure}
 
\section{Constraints on anomalous Higgs boson interactions}
 
The strength and tensor structure of the Higgs boson interactions
are investigated using an effective Lagrangian
approach~\cite{ATL-PHYS-PUB-2017-018}.
Extra terms of the form $c_i^{(D)}\mathcal{O}_i^{(D)}/\Lambda^{D-4}$,
where $\Lambda$ is the energy scale of the new interactions,
${O}_i^{(D)}$ are dimension-$D$ operators, and
$c_i^{(D)}$ are numerical coefficients,
are added to the SM Lagrangian to obtain
an effective Lagrangian inspired by that in Ref.~\cite{Contino:2013kra}.
Only dimension $D=6$ operators are considered in this study,
since dimension $D=5$ operators violate lepton or baryon number,
while dimension $D>6$ operators are further suppressed by powers of
$\Lambda$.
 
The results presented in this \type\ focus on the coefficients of
the operators in the `Strongly Interacting Light Higgs'
formulation~\cite{Giudice:2007fh}. This formalism is
defined as the effective theory of a strongly interacting
sector in which a  light composite Higgs boson arises as a pseudo Goldstone
boson, and is responsible for EW symmetry breaking.
Among such operators, four directly affect the $VH$ cross-sections
because they introduce new Higgs boson interactions with $W$ bosons
($\mathcal{O}_{HW}$, $\mathcal{O}_{W}$) and $Z$ bosons (all four operators):
\begin{itemize}
\item $\mathcal{O}_{HW} = i\left(D^\mu H\right)^\dagger\sigma^a\left(D^\nu H\right)W^a_{\mu\nu}$,
\item $\mathcal{O}_{HB} = i\left(D^\mu H\right)^\dagger\left(D^\nu H\right)B_{\mu\nu}$,
\item $\mathcal{O}_{W} = \frac{i}{2}\left(H^\dagger\sigma^a\overset{\text{\scriptsize$\leftrightarrow$}}{D^\mu}H\right)D^\nu W^a_{\mu\nu}$,
\item $\mathcal{O}_{B} = \frac{i}{2}\left(H^\dagger \overset{\text{\scriptsize$\leftrightarrow$}}{D^\mu} H\right)\partial^\nu B_{\mu\nu}$.
\end{itemize}
 
The corresponding $CP$-odd operators $\tilde{\mathcal{O}}_{HW}$,
$\tilde{\mathcal{O}}_{HB}$, $\tilde{\mathcal{O}}_{W}$, and
$\tilde{\mathcal{O}}_{B}$, are not considered.
 
Modifications of the $gg\to ZH$ production cross-section are only introduced
by either higher-dimension ($D \geq 8$) operators or corrections
that are formally at NNLO in QCD, and are
not included in this study, in which the expected $gg\to ZH$
contribution is kept fixed to the SM prediction.
 
The operator $\mathcal{O}_d = y_d|H|^2\bar{Q}_\mathrm{L}Hd_\mathrm{R}$
(plus Hermitian conjugate) with Yukawa coupling strength $y_d$,
which modifies the coupling between the Higgs boson
and down-type quarks, induces variations of the partial width $\Gamma_H^{bb}$
and of the total Higgs boson width $\Gamma_H$,
and therefore of the $H\to b\bar{b}$ branching ratio. This operator affects
the measured cross-sections in the same way in each
region.
 
Constraints are set on the coefficients of the five
$\mathcal{O}_{W}$, $\mathcal{O}_B$, $\mathcal{O}_{HW}$,
$\mathcal{O}_{HB}$ and $\mathcal{O}_d$
operators in the `Higgs Effective Lagrangian' (HEL)
implementation~\cite{Alloul:2013naa}, using the known
relations between such coefficients and the stage-1 STXS
based on leading-order predictions~\cite{Hays:2290628}.
Such relations include interference terms between the
SM and non-SM amplitudes that are linear in the coefficients
and of order $1/\Lambda^2$, and the SM-independent contributions
that are quadratic in the coefficients and of order $1/\Lambda^4$.
In the HEL implementation, the coefficients $c_i$ of interest
are recast into the following dimensionless coefficients:
\begin{equation*}
\bar{c}_{HW}=\frac{m_W^2}{g}\frac{c_{HW}}{\Lambda^2},\ \ \
\bar{c}_{HB}=\frac{m_W^2}{g^\prime}\frac{c_{HB}}{\Lambda^2},\ \ \
\bar{c}_{W}=\frac{m_W^2}{g}\frac{c_{W}}{\Lambda^2},\ \ \
\bar{c}_{B}=\frac{m_W^2}{g^\prime}\frac{c_{B}}{\Lambda^2},\ \ \
\bar{c}_d=v^2 \frac{c_{d}}{\Lambda^2},
\end{equation*}
where $g$ and $g^\prime$ are the $SU(2)$ and $U(1)$ SM gauge couplings,
and $v$ is the vacuum expectation value of the Higgs boson field.
These dimensionless coefficients are equal to zero in the SM.
 
The sum $\bar{c}_{W}+\bar{c}_{B}$ is strongly constrained
by precision EW data~\cite{Ellis:2018gqa}
and is thus assumed here to be zero,
and constraints are set on $\bar{c}_{HW}$,  $\bar{c}_{HB}$,
$\bar{c}_{W}-\bar{c}_{B}$ and $\bar{c}_d$.
The relations between the HEL coefficients and the reduced STXS
measured in this \type\ are obtained by averaging the relations
for the regions that are merged with weights proportional to their
respective cross-sections.
 
Simultaneous maximum-likelihood fits to the five STXS
measured in the 5-POI scheme are performed
to determine $\bar{c}_{HW}$, $\bar{c}_{HB}$,
$\bar{c}_{W}-\bar{c}_{B}$ and $\bar{c}_d$.
Due to the large sensitivity to the Higgs boson anomalous
couplings to vector bosons provided by the $\pT^V>250$~\GeV\ cross-sections,
the 5-POI results place tighter constraints on these
coefficients (e.g. approximately a factor two for $\bar{c}_{HW}$)
than do the 3-POI results. For this reason, constraints
obtained with the 3-POI results are not shown here.
 
In each fit, all coefficients but one are assumed to vanish,
and 68\% and 95\% confidence level (CL) one-dimensional intervals are
inferred for the remaining coefficient.
The negative-log-likelihood one-dimensional projections
are shown in Figure~\ref{fig:EFT_1Dscans}, and the
68\% and 95\% CL intervals for $\bar{c}_{HW}$, $\bar{c}_{HB}$,
$\bar{c}_{W}-\bar{c}_{B}$ and $\bar{c}_d$
are summarised in Table~\ref{tab:EFT_results_1D}.
The parameters $\bar{c}_{HW}$ and $\bar{c}_{W}-\bar{c}_{B}$ are constrained
at 95\%~CL to be no more than a few percent, while the constraint on
$\bar{c}_{HB}$ is about five times worse,
and the constraint on $\bar{c}_d$ is of order unity.
For comparison, Table~\ref{tab:EFT_results_1D} also
shows the 68\% and 95\% CL intervals
for the dimensionless coefficients when
the SM-independent contributions, which are
of the same order ($1/\Lambda^4$) as the
dimension-8 operators that are neglected,
are not considered. The constraints are typically
50\% stronger than when
the SM-independent contributions are not neglected.

\begin{figure}[!ht]
\centering
\subfloat[]{\includegraphics[width=0.5\linewidth]{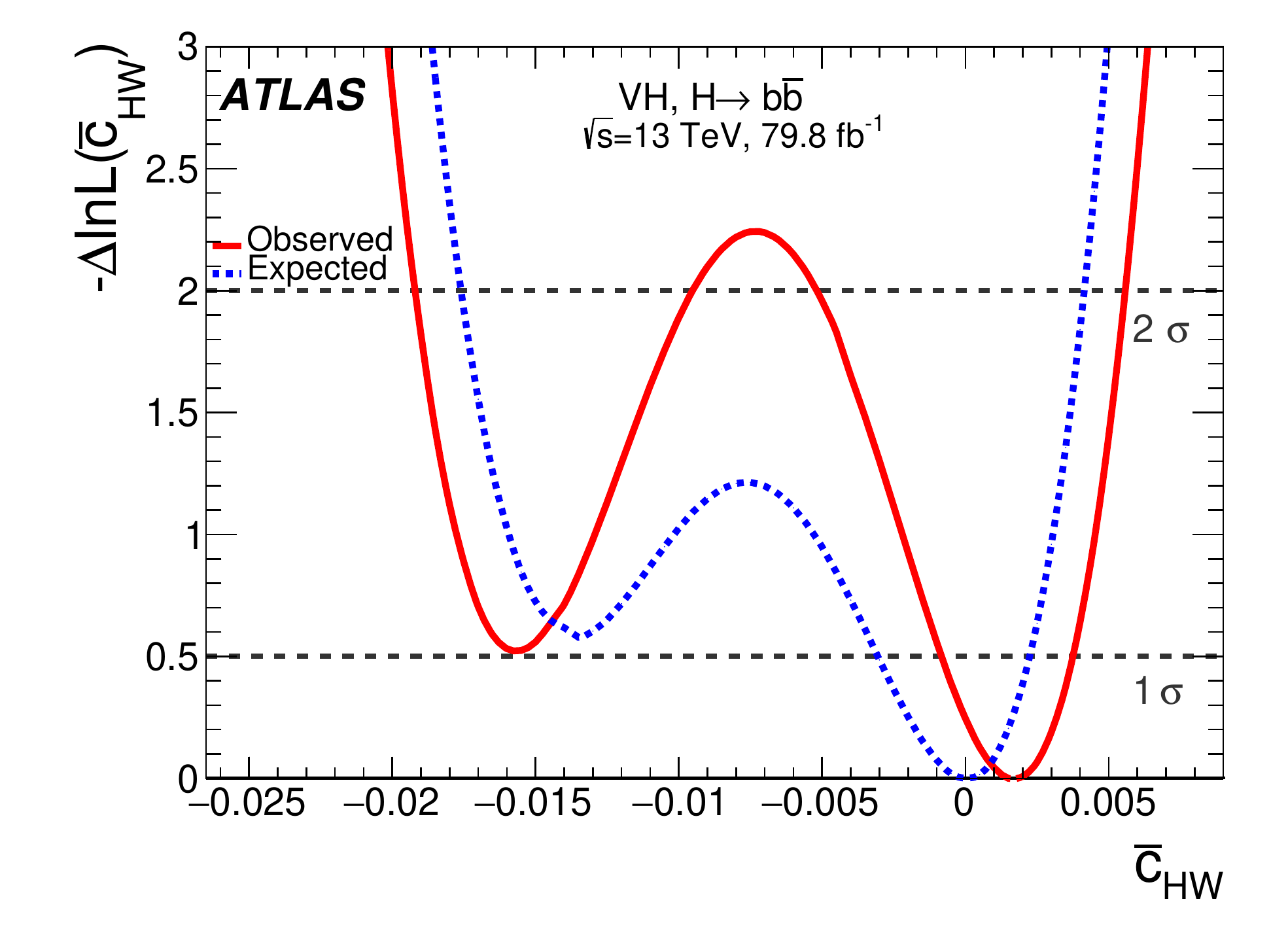}}
\subfloat[]{\includegraphics[width=0.5\linewidth]{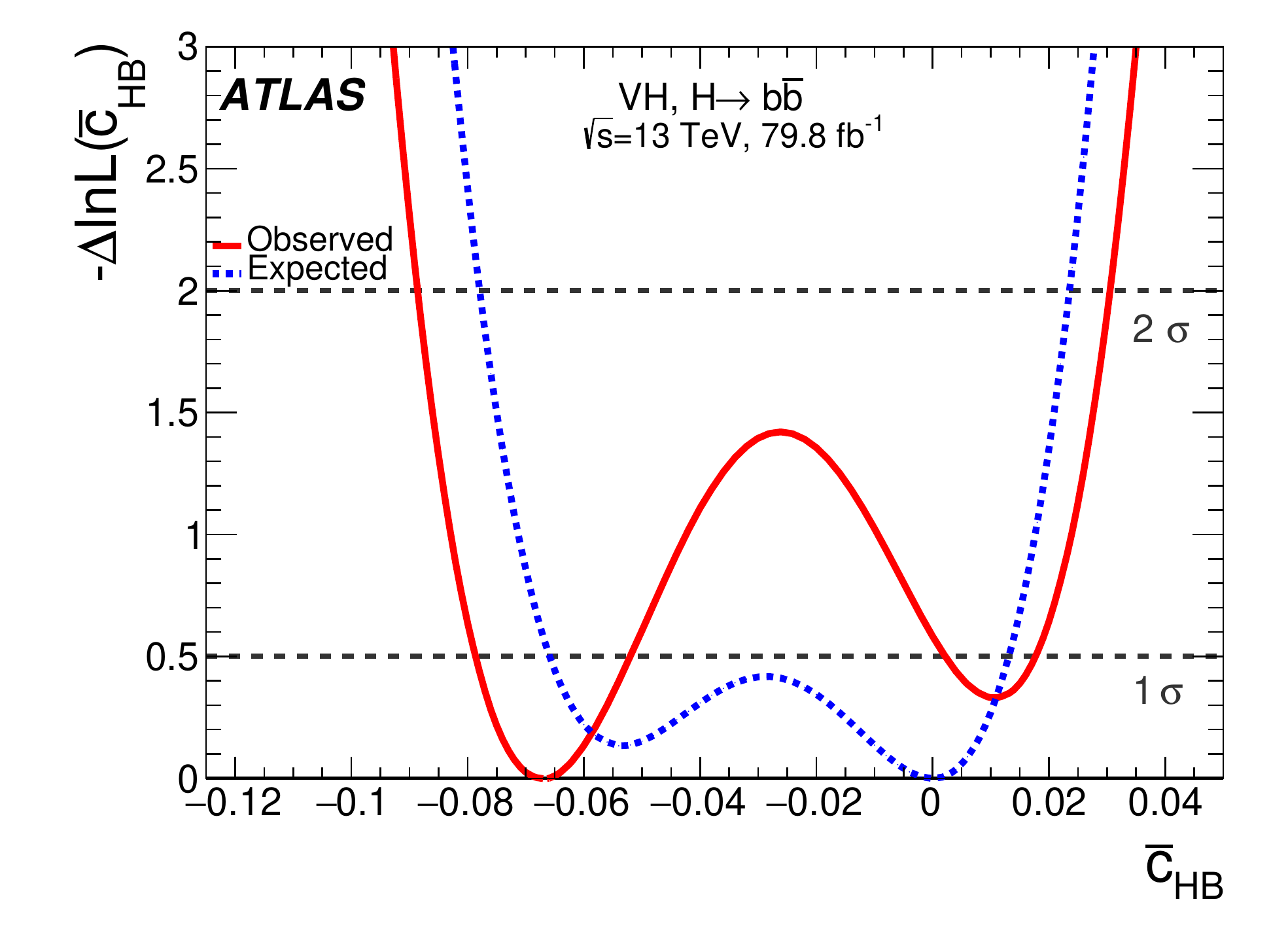}}\\
\subfloat[]{\includegraphics[width=0.5\linewidth]{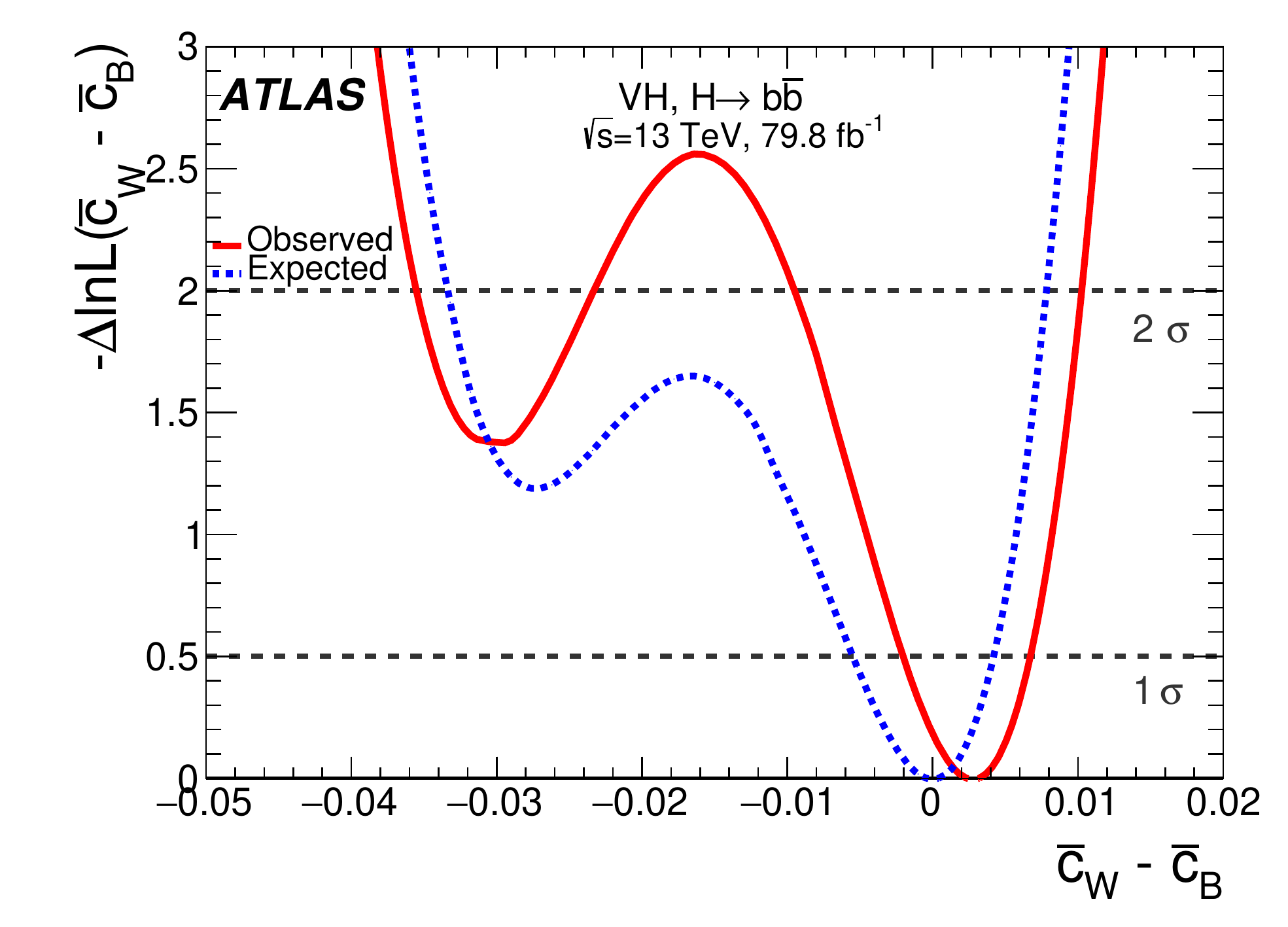}}
\subfloat[]{\includegraphics[width=0.5\linewidth]{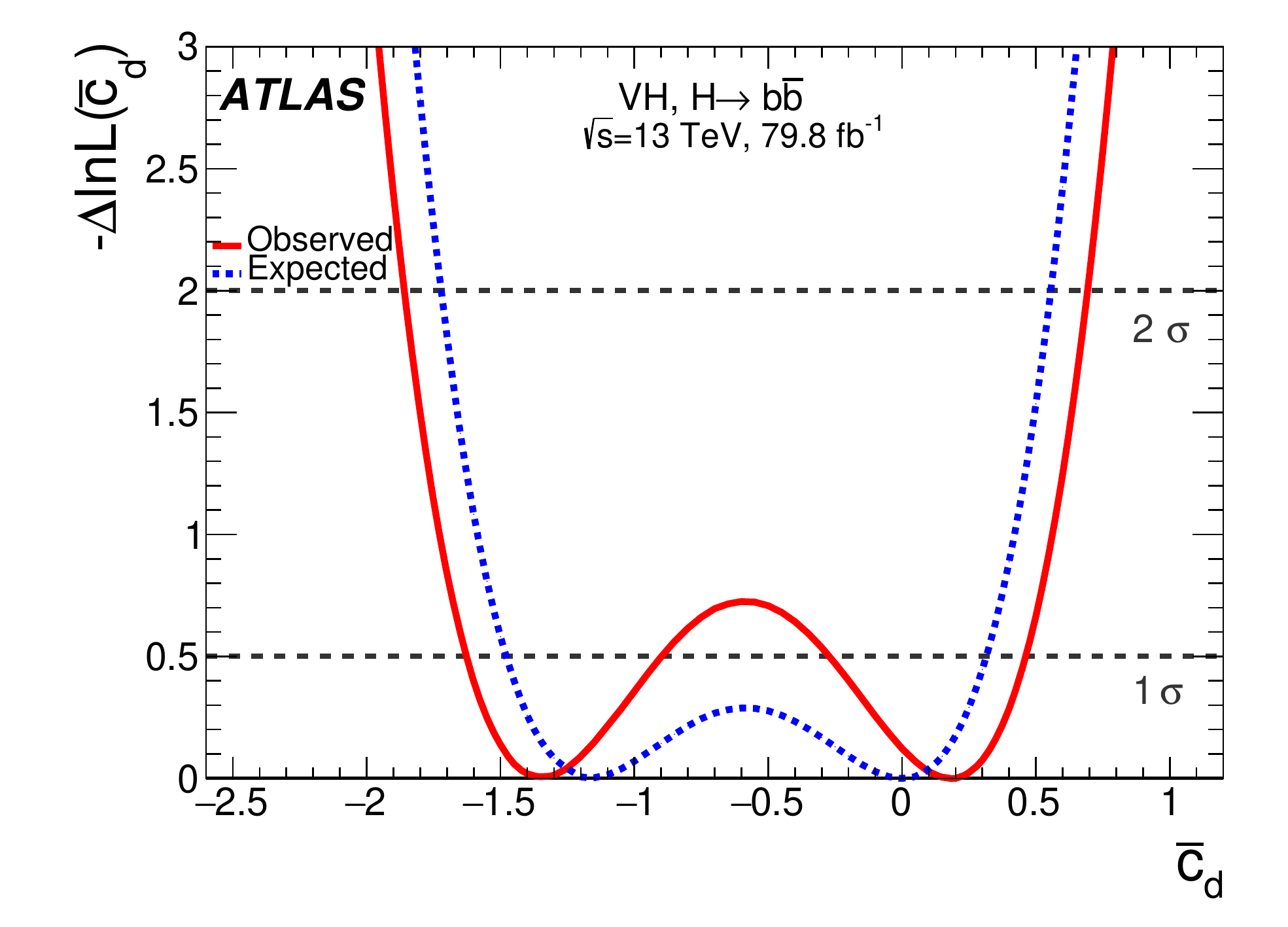}}
\caption{The observed (solid) and expected (dotted) profiled
negative-log-likelihood functions for the one-dimensional
fits to constrain the coefficients (a) $\bar{c}_{HW}$,
(b) $\bar{c}_{HB}$, (c) $\bar{c}_{W}-\bar{c}_{B}$
and (d) $\bar{c}_{d}$
of an effective Lagrangian (described in the text),
when the other coefficients are assumed to vanish.}
\label{fig:EFT_1Dscans}
\end{figure}

\begin{table}[!htbp]
\begin{center}
\caption{The expected and observed 68$\%$ CL (four top rows) and 95\% CL
(four bottom rows) intervals for the
effective Lagrangian coefficients
$\bar{c}_{HW}$, $\bar{c}_{HB}$, $\bar{c}_{W}-\bar{c}_{B}$
and $\bar{c}_d$
when the other coefficients are assumed to vanish.
Each row is composed of two sub-rows: the first one uses the
interference between SM and non-SM amplitudes and the SM-independent
contributions, while the second sub-row uses only the
interference between SM and non-SM amplitudes.
}
\label{tab:EFT_results_1D}
\begin{tabular}{lcc}
\hline
\hline
Coefficient         & Expected interval                                & Observed interval   \\
\hline
\hline
\multicolumn{3}{c}{Results at 68\% confidence level} \\
\hline
$\bar{c}_{HW}$             & [$-0.003$, 0.002]                                & [$-0.001$, 0.004]\\
~~(interference only       & [$-0.002$, 0.003]                                &  [$-0.001$, 0.005])\\\hdashline
$\bar{c}_{HB}$             & [$-0.066$, 0.013]                                 & [$-0.078$, $-0.055$] $\bigcup$ [0.005,  0.019]\\
~~(interference only       & [$-0.016$, 0.016]                                & [$-0.005$, 0.030])\\\hdashline
$\bar{c}_{W}-\bar{c}_{B}$  & [$-0.006$, 0.005]                                 & [$-0.002$, 0.007]\\
~~(interference only       & [$-0.005$, 0.005] &  [$-0.002$, 0.008])\\\hdashline
$\bar{c}_d$               &  [$-1.5$, 0.3]                            & [$-1.6$, $-0.9$] $\bigcup$ [$-0.3$, $0.4$] \\
~~(interference only & [$-0.4$, 0.4] & [$-0.2$, $0.7$]) \\
\hline
\multicolumn{3}{c}{Results at 95\% confidence level} \\
\hline
$\bar{c}_{HW}$            & [$-0.018$, 0.004]              & [$-0.019$,$-0.010$] $\bigcup$ [$-0.005$, 0.006] \\
~~(interference only & [$-0.005$, 0.005] & [$-0.003$, $0.008$])\\\hdashline
$\bar{c}_{HB}$            & [$-0.078$, 0.024]                & [$-0.090$, $0.032$]                             \\
~~(interference only & [$-0.033$, 0.033] & [$-0.022$, $0.049$])\\\hdashline
$\bar{c}_{W}-\bar{c}_{B}$ & [$-0.034$, 0.008]          & [$-0.036$,$-0.024$] $\bigcup$ [$-0.009$, 0.010] \\
~~(interference only & [$-0.009$, 0.010] & [$-0.006$, $0.014$])\\\hdashline
$\bar{c}_d$              & [$-1.7$, $0.5$]                                  & [$-1.9$, $0.7$] \\
~~(interference only & [$-0.8$, 0.8] & [$-0.6$, $1.1$])\\
\hline
\hline
\end{tabular}
\end{center}
\end{table}
 
 
\FloatBarrier
 
\section{Conclusion}
Using 79.8~\ifb\ of $\sqrt{s}=13$~\TeV\ proton--proton collisions
collected by the ATLAS detector at the LHC,
the cross-sections for the associated production of a Higgs boson decaying
into bottom-quark pairs and an electroweak gauge boson $W$ or $Z$
decaying into leptons are measured as functions of the vector-boson
transverse momentum $\pT^V$.
The cross-sections are measured for Higgs bosons in a fiducial volume with
rapidity  $|y_H|<2.5$, in the `simplified template cross-section' framework.
 
The measurements are performed for two different choices of the
number of $\pT^V$ intervals.
The results have relative uncertainties varying
between 50\% and 125\% in one case,
and between 29\% and 56\% in the other.
The measurements are in agreement with the Standard Model
predictions, even in high $\pT^V$ ($>250$~\GeV) regions that are most
sensitive to enhancements from potential anomalous interactions between
the Higgs boson and the electroweak gauge bosons.
 
One-dimensional limits on four linear combinations
of the coefficients of effective Lagrangian operators
affecting the Higgs boson couplings to the electroweak gauge bosons
and to down-type quarks
have also been set.
For two of these parameters the constraint has a precision of
a few percent.
 
\section*{Acknowledgements}
 
We thank CERN for the very successful operation of the LHC, as well as the
support staff from our institutions without whom ATLAS could not be
operated efficiently.
 
We acknowledge the support of ANPCyT, Argentina; YerPhI, Armenia; ARC, Australia; BMWFW and FWF, Austria; ANAS, Azerbaijan; SSTC, Belarus; CNPq and FAPESP, Brazil; NSERC, NRC and CFI, Canada; CERN; CONICYT, Chile; CAS, MOST and NSFC, China; COLCIENCIAS, Colombia; MSMT CR, MPO CR and VSC CR, Czech Republic; DNRF and DNSRC, Denmark; IN2P3-CNRS, CEA-DRF/IRFU, France; SRNSFG, Georgia; BMBF, HGF, and MPG, Germany; GSRT, Greece; RGC, Hong Kong SAR, China; ISF and Benoziyo Center, Israel; INFN, Italy; MEXT and JSPS, Japan; CNRST, Morocco; NWO, Netherlands; RCN, Norway; MNiSW and NCN, Poland; FCT, Portugal; MNE/IFA, Romania; MES of Russia and NRC KI, Russian Federation; JINR; MESTD, Serbia; MSSR, Slovakia; ARRS and MIZ\v{S}, Slovenia; DST/NRF, South Africa; MINECO, Spain; SRC and Wallenberg Foundation, Sweden; SERI, SNSF and Cantons of Bern and Geneva, Switzerland; MOST, Taiwan; TAEK, Turkey; STFC, United Kingdom; DOE and NSF, United States of America. In addition, individual groups and members have received support from BCKDF, CANARIE, CRC and Compute Canada, Canada; COST, ERC, ERDF, Horizon 2020, and Marie Sk{\l}odowska-Curie Actions, European Union; Investissements d' Avenir Labex and Idex, ANR, France; DFG and AvH Foundation, Germany; Herakleitos, Thales and Aristeia programmes co-financed by EU-ESF and the Greek NSRF, Greece; BSF-NSF and GIF, Israel; CERCA Programme Generalitat de Catalunya, Spain; The Royal Society and Leverhulme Trust, United Kingdom.
 
The crucial computing support from all WLCG partners is acknowledged gratefully, in particular from CERN, the ATLAS Tier-1 facilities at TRIUMF (Canada), NDGF (Denmark, Norway, Sweden), CC-IN2P3 (France), KIT/GridKA (Germany), INFN-CNAF (Italy), NL-T1 (Netherlands), PIC (Spain), ASGC (Taiwan), RAL (UK) and BNL (USA), the Tier-2 facilities worldwide and large non-WLCG resource providers. Major contributors of computing resources are listed in Ref.~\cite{ATL-GEN-PUB-2016-002}.
 

\printbibliography

\clearpage
 
\begin{flushleft}
{\Large The ATLAS Collaboration}

\bigskip

M.~Aaboud$^\textrm{\scriptsize 35d}$,    
G.~Aad$^\textrm{\scriptsize 101}$,    
B.~Abbott$^\textrm{\scriptsize 128}$,    
D.C.~Abbott$^\textrm{\scriptsize 102}$,    
O.~Abdinov$^\textrm{\scriptsize 13,*}$,    
A.~Abed~Abud$^\textrm{\scriptsize 70a,70b}$,    
D.K.~Abhayasinghe$^\textrm{\scriptsize 93}$,    
S.H.~Abidi$^\textrm{\scriptsize 167}$,    
O.S.~AbouZeid$^\textrm{\scriptsize 40}$,    
N.L.~Abraham$^\textrm{\scriptsize 156}$
% ...
\bigskip
\\

% ...
$^{13}$Institute of Physics, Azerbaijan Academy of Sciences, Baku; Azerbaijan.\\
% ...
$^{35}$$^{(a)}$Facult\'e des Sciences Ain Chock, R\'eseau Universitaire de Physique des Hautes Energies - Universit\'e Hassan II, Casablanca;$^{(b)}$Facult\'{e} des Sciences, Universit\'{e} Ibn-Tofail, K\'{e}nitra;$^{(c)}$Facult\'e des Sciences Semlalia, Universit\'e Cadi Ayyad, LPHEA-Marrakech;$^{(d)}$Facult\'e des Sciences, Universit\'e Mohamed Premier and LPTPM, Oujda;$^{(e)}$Facult\'e des sciences, Universit\'e Mohammed V, Rabat; Morocco.\\
% ...
$^{40}$Niels Bohr Institute, University of Copenhagen, Copenhagen; Denmark.\\
% ...
$^{70}$$^{(a)}$INFN Sezione di Pavia;$^{(b)}$Dipartimento di Fisica, Universit\`a di Pavia, Pavia; Italy.\\
% ...
$^{93}$Department of Physics, Royal Holloway University of London, Egham; United Kingdom.\\
% ...
$^{101}$CPPM, Aix-Marseille Universit\'e, CNRS/IN2P3, Marseille; France.\\
$^{102}$Department of Physics, University of Massachusetts, Amherst MA; United States of America.\\
% ...
$^{128}$Homer L. Dodge Department of Physics and Astronomy, University of Oklahoma, Norman OK; United States of America.\\
% ...
$^{156}$Department of Physics and Astronomy, University of Sussex, Brighton; United Kingdom.\\
% ...
$^{167}$Department of Physics, University of Toronto, Toronto ON; Canada.\\
% ...
$^{a}$ Also at Centre for High Performance Computing, CSIR Campus, Rosebank, Cape Town; South Africa.\\
$^{b}$ Also at CERN, Geneva; Switzerland.\\
$^{c}$ Also at CPPM, Aix-Marseille Universit\'e, CNRS/IN2P3, Marseille; France.\\
$^{d}$ Also at D\'epartement de Physique Nucl\'eaire et Corpusculaire, Universit\'e de Gen\`eve, Gen\`eve; Switzerland.\\
$^{e}$ Also at Departament de Fisica de la Universitat Autonoma de Barcelona, Barcelona; Spain.\\
% ...
$^{*}$ Deceased

\end{flushleft}

\end{document}